# Not Just a Matter of Time:
# Field Differences and the Shaping of Electronic Media in Supporting Scientific Communication

**Rob Kling & Geoffrey McKim**

April 27, 2000

Indiana University School of Library and Information Science
10th & Jordan, Bloomington, IN 47405 USA
+1 812 855 5113
kling@indiana.edu, mckimg@indiana.edu

## Abstract

The shift towards the use of electronic media in scholarly communication appears to be an inescapable imperative. However, these shifts are uneven, both with respect to field and with respect to the form of communication. Different scientific fields have developed and use distinctly different communicative forums, both in the paper and electronic arenas, and these forums play different communicative roles within the field. One common claim is that we are in the early stages of an electronic revolution, that it is only a matter of time before other fields catch up with the early adopters, and that all fields converge on a stable set of electronic forums. A social shaping of technology (SST) perspective helps us to identify important social forces – centered around disciplinary constructions of trust and of legitimate communication – that pull against convergence. This analysis concludes that communicative plurality and communicative heterogeneity are durable features of the scholarly landscape, and that we are likely to see field differences in the use of and meaning ascribed to communications forums persist, even as overall use of electronic communications technologies both in science and in society as a whole increases.

## Introduction

The use of electronic media to support scientific communication is one of the major shifts in the practice of science in this era. There are other shifts in the science system, such as the rise of global science, the increasing importance of the biomedical sciences, the plateauing of support for mega-science projects after the end of the Cold War. There are interdependencies in these shifts – since electronic communication media can often expedite special kinds of communications between scientists who work across continents and 10-15 time zones while reducing the marginal costs of communication.

Today, the Internet is the primary medium of this communication. In North America, public access to the Internet has become the occasion for both discourse about and changes in ways of doing business, forms of entertainment, communication within families, and so on. As a consequence, the shift towards using electronic media as a major communication medium seems to be an inescapable imperative. The concept of an inescapable imperative has not become popular as a finding of scientific research; rather, it is popular because it fits simple cultural models of computerization and because it is advanced in many important forums. We do not agree with this view, and will present our arguments in this article.

It is easy to give enough examples of diverse practices such as the communication of conference programs as they jell, the sharing of preprints, access to electronic versions of journal articles (both before and after paper publication), and the development of shared disciplinary corpuses so that they appear to be sweeping across the sciences. However, each of these practices seems to be emblematic of specific fields rather than developing in ways that will make them universal.

It is also easy to be sanguine about this differential pattern of developments. One argument is that "sooner or later everyone will catch on" and learn to use the various e-media structures in all fields. It is just a matter of time before laggard fields catch up with the leaders, and the scholarly community converges on a stable set of electronic forums, such as "pre-print" servers, discussion lists, and electronic journals. A second common argument is that the variety of e-media initiatives reflects a creative period in scientific communication, and this, in itself, is a good thing. Finally, some science policy analysts suggest that we actively identify "best practices" and encourage their widespread adoption.

We see notable risks in a pure laissez-faire "let them work it out for themselves" approach. Large amounts of money, resources, and effort are being committed (by government agencies, by academic departments, by publishers, by professional societies, and by individual researchers) to the development, maintenance, and promotion of various forms of communications technologies for use in global science. However, in the absence of a valid theory of how scholarly fields adopt and shape technology, scientists and policy-makers are left only with context-free models, and hence resources may be committed to projects that are not self-sustainable, that wither, and that do not effectively improve the scientific communications system of the field. The consequences may not only be sub-optimal use of financial resources, but also wasted effort on the part of individual researchers, and even data languishing in marginal, decaying, and dead systems and formats.

The purpose of this *JASIS* perspectives article is to deepen our understanding of the future of electronic communications in science. It is difficult to predict the long-term future, as too many contingencies can and will shape long-term outcomes to make meaningful predictions. One may casually predict that that many if not all research journals will issue electronic editions in the 21st century; however, more complex issues about the role of paper editions, the extent and nature of peer review, and the journals' pricing/revenue models are hard to casually predict for well over a century away.

However, we can be more precise about short- to medium-term (up to around 20 years, or approximately a generation) shifts in science communications if we identify some of the key social forces. After all, many of the faculty who have established productive work and communication practices and who just received tenure this year will still be active in some aspects of creating or reviewing research articles 20 years from now. Further, Covi (in press) found that apprentice scholars (such as graduate students and post-docs) often emulated their faculty advisors' stances towards the value and use of specific electronic publication sources and archives. The range of technological possibilities may change rapidly, while the worldviews that dominate specific scientific communities are likely to change much more slowly.

This article begins by examining **heterogeneity** of communications practices among fields, examining field differences as an opportunity to understand the social forces that shape the use of, and shifts in the use of, electronic media in scientific communication. The dominant analyses of the role of electronic media in scholarly communication focus on the information processing costs and speeds of different media. These "information processing" analyses are field-independent: the costs and speed of publishing in paper or electronic media in two fields, such as chemistry and physics should be similar (except for differences in production expenses for artwork or color). Information processing analyses lead one to predict that differences in communication practices across fields should diminish over time. Hars (1999a) develops an information processing analysis of corpuses of scientific preprints. He comments:

> "my argument is generic and should apply to any scientific discipline. Thus I expect (the fields of) Information Systems and Chemistry to embrace online publishing in a similar way as physics, etc. It may just take them longer. (Certainly, this is a naive view and there may be factors rooted in power, tradition etc. which may hinder this development. But I don't think that there is a systematic difference between chemistry/Information Systems and other disciplines which prevents the former from adopting similar structures for online knowledge infrastructures.) (Hars, 1999b).

Similarly, Paul Ginsparg, the developer of the Los Alamos Physics E-Print Server, writes:

> "Regardless of how different research areas move into the future (perhaps by some parallel and ultimately convergent evolutionary paths), I strongly suspect that on the one- to two-decade time scale, serious research biologists will also have moved to some form of global unified archive system, without the current partitioning and access restrictions familiar from the paper medium, for the simple reason that it is the best way to communicate knowledge, and hence to create new knowledge (Ginsparg, 1999)."

In contrast, we examine scholarly communication systems as embedded in work organizations that produce the analyses which subsequently appear in venues such as seminars, conferences talks, articles, reports, and books. The organization of work and of access to resources to conduct research may influence the conventions about legitimate forms of scholarly communication. Communication conventions change over time. But, in this "institutional view" they are embedded in ways of organizing work that differ between fields and that may change slowly (Kling & Iacono, 1989). In this institutional view, heterogeneous communication practices are durable in the medium-term. This analysis leads us to believe that *it is not just a matter of time* for disciplines to converge on common ways to using e-media to support scholarly communication.

## Field Differences in Shaping of Electronic Media

One of the notable features about the development of e-media in science is that they seem to vary in their structure, roles and uses from one field (or closely related set of fields) to another[1]. This observation has been advanced most extensively in Crawford, Hurd, & Weller (1996), who provide detailed case studies of the communications practices in the human genetics, high-energy physics, and space sciences fields. This observation is robust with respect to the ways that scientists use e-mail and collections of paper and electronic journals (see Kling & Covi, 1995; Walsh & Bayma, 1996; Finholt & Brooks, 1997; Kling & Covi 1997; Walsh & Bayma, 1997; Covi & Kling, 1998; Walsh, 1997).

The contrast between the organization and roles of electronic materials in several sciences is instructive and is one starting point for analysis. While every field will use electronic media slightly differently (for different purposes, in different forums, at different rates), we will examine three exemplar fields – high-energy physics, molecular biology, and information systems -- that illustrate distinctive configurations of communications forums.

### High Energy Physics

The field of experimental high-energy physics is dominated by a relatively small number of large, highly-visible projects. High-energy physics projects frequently have long lifetimes – projects lasting 3-5 years or even longer are typical (Traweek, 1992). Major high-energy physics experiments are carried out by large, multi-institutional collaborations, sometimes with budgets in the hundreds of millions of dollars. For example, the D0, CDF, and ALEPH collaborations all had upwards of 400 members each, from over 30 institutions each. The primary equipment required for high-energy physics experiments – colliding beam detectors – are extremely expensive, and therefore highly centralized in a few major laboratories. In the mid 1990s, there were only 6 colliding detectors in the world available to physicists (Gallison, 1997).

High-energy physicists have long lead the sciences in the use of electronic media for sharing working papers. For example, most high-energy physicists contribute draft articles to an electronic working paper server at the time of submission of the article to a paper journal. While the paper journals are still important for archiving and for prestige and reward allocation, these electronic working paper servers are frequently the primary means of formal communication. The working paper (or "e-print") server at Los Alamos National Labs has become central in the communications system of the field (Odlyzko 1996). While the Los Alamos National Labs E-Print server is the best-known of these electronic working paper servers in the U.S., it isn't the only one. There are about 11 others, (including the CERN preprint server at http://preprints.cern.ch/, DESY preprints at ftp://ftp.desy.de/pub/preprints/, and the American Physical Society at http://publish.aps.org/eprint/).

The current centrality of the Web, in the form of these e-print servers, in high-energy physics is frequently attributed to the central role of high-energy physicists (most important, Tim Berners-Lee, from CERN) in having developed the basic protocols underlying the World Wide Web. However, the use of electronic working paper servers in high-energy physics predated the Web; early e-print servers used FTP (file transfer protocol) and electronic mail to collect and make available electronic working papers (Kreitz, et al., 1996). Even more importantly, high-energy physicists had a pre-print culture that long predated their use of electronic media. Physicists since the 1970s submitted articles to paper-based pre-print clearinghouses, that then redistributed the papers to interested researchers (Kreitz, et al, 1996).

### Molecular Biology

In contrast with high-energy physicists, molecular biologists typically circulate preprints only within small invisible colleges and broader access depends upon publication in archival journals. While a few electronic working-paper servers have been established in a few biology subfields[2], such servers do not play a noticeable role in the communications system of the field. Many biological fields, however, do use digital disciplinary corpora or other shared databases for important and growing data sets. The Protein Data Bank (PDB) is used as a repository for experimentally determined three-dimensional structures of biological macromolecules by researchers in the protein sciences. FlyBase (http://flybase.bio.indiana.edu/) provides *Drosophila* (fruit fly) researchers with access to submitted genomic data, genetic maps of *Drosophila*, addresses of other *Drosophila* researchers, fly stock lists (directories of which labs supply which genetic varieties of fruit fly), and a bibliography of publications on *Drosophila*. The *C. elegans* community, which studies Nematode worms, uses ACEDB (A C. Elegans Data Base). Other model-organism-oriented biology subfields have similar digital disciplinary corpora.

Submission of gene sequences or other experimentally determined data into these shared databases when appropriate is required before publication in many molecular biology journals. The "accession number", a unique number identifying a dataset submitted to one of these databases, is then published along with an article in a paper journal, allowing readers to obtain research data almost instantly, if desired. Such digital disciplinary corpora have become critical to the communications systems of the fields. However, unlike in high-energy physics, these

---

biological databases operate synergistically, rather than competitively, with paper journals.

## Information Systems

Information systems researchers are specially interested in strategies for deciding which organizational activities should be computerized (and how), developing effective information systems in organizations, for evaluating information systems, and for understanding how people use systems in the context of organizational change processes. Information systems is a relatively new field, which was formed in the 1970s and has primarily been located in schools of management or business. It had roots in management science, data processing, accounting, and organizational behavior. In the 1980s the field jelled rapidly with its own specialty journals, research conferences, and Ph.D. programs.

The information systems discipline has developed ISWORLD, an extensive Web-based collection of links, articles, course syllabi, tools, and resources that are maintained in a distributed fashion (http://www.isworld.org/isworld/isworldtext.html). Eight scholarly societies are listed as co-founders and the top-level Web site is sponsored by *MIS Quarterly*, the top-impact journal in the field. Information Systems scholars act as section editors for the many sub-pages of the site and the result is in an extensive, distributed but centrally accessible digital disciplinary corpus. The research sections contain links to tutorials, software, field overviews, and so on (http://www.isworld.org/ isworld/isworldtext.html#research). However, unlike the E-Print Server at LANL, ISWORLD is not a repository for the full text of working papers that have yet to be refereed for conferences or journals. However, a small fraction of information systems faculty do post some of their working papers on their own web sites or in the report series of a research center at their universities. ISWORLD also manages several associated electronic discussion lists which information systems scholars use for making inquiries about research topics and teaching materials, as well as conference announcements and other publishing opportunities.,

Table 1 illustrates in tabular for some the heterogeneity found in the use of electronic media in six different fields, by enumerating key electronic forums in these fields. These forums are each briefly described in Appendix A.

**Table 1: Examples of Electronic Communications Forums by Type and Field**

| Field | E-Print Servers | Pure Electronic Journals | Digital Disciplinary Corpora | Paper-Electronic Journal Enhancements | Shared Digital Libraries |
|---|---|---|---|---|---|
| High-energy physics | SLAC-SPIRES, XXX/Arxiv.org | | | *Science Online* | |
| Computer Science | NCESTRL | *Chicago Journal of Theoretical Computer Science* | | *ACM* journals, CS journals published by Elsevier, Springer-Verlag | ACM Digital Library |
| Artificial Intelligence | | *ETAI; J. of AI Research (JAIR)* | | *Artificial Intelligence Journal* | |
| Molecular biology | | | FlyBase, GDB, GenBank, PDB | *Science Online* | PubMed Central |
| Astrophysics | | | Hubble Space Telescope (HST) Archive | *Science Online; The Astrophysical Journal* | Astrophysics Data System (ADS) |
| Information systems | | *J. Assoc. Info Systems; MIS Discovery* | ISWORLD | *MIS Quarterly* | ISWORLD |

## Social Forces and the Shaping of E-Media: Convergence or Heterogeneity

In this section, we will discuss a series of social forces that our own research suggests play a role in the shaping of electronic media in scholarly communication, leading towards convergence and towards increasing heterogeneity in some of their key features.

## The Electronic Publishing  Reform Movement

Electronic communication is not simply a set of professional practices; it is also the focus of a small e-publishing professional reform movement.  This reform movement shares much in common with other computerization movements (Kling & Iacono, 1995; Iacono & Kling, 1996).

First, it is energized and most widely publicized by a *core group of enthusiasts* (e.g. Paul Ginsparg, Stevan Harnad, and Andrew Odlyzko).  Harnad is well-known as the editor of the electronic  journal *Psycholoquy*, as the originator of "scholarly skywriting", a short, discursive, and iterative form of scholarly communication (Harnad, 1991), and for his "subversive proposal", a radically decentralized scholarly publishing model, in which scholars self-publish their works, which then may or may not be peer-reviewed (Brent, 1995).  Ginsparg is best known as the developer of the Los Alamos National Labs Physics E-Print Server, a working paper server used by high-energy physicists (http://www.arxiv.org/).  Morton (1997) urges scholars to accept the "paradigm shift" and switch to (and form, if necessary), electronic communication, in both centralized and decentralized forms.

Second, the movement's activists advocate a *shared ideology*.  This ideology has as its primary precept that "electronic media are almost always better than traditional media, such as paper", for several related reasons: electronic communication is dramatically less expensive than alternatives; access to electronic media is easier and wider; and the systematic use of electronic media can speed up scientific communication[3].

Many of the strongest claims, both for the **inevitability** of widespread, or even universal, use of electronic media, and, more specifically, **convergence** on a set of e-media forms come out of the reform movement. Claims that the near-universal adoption of electronic media is inevitable are generally predicated on two putative attributes of electronic media: their radically different (and lower) cost structures, and their new technological features. Odlyzko articulates the inevitability argument most clearly:

> "I expect that scholarly publishing will move to almost exclusively electronic means of information dissemination.   This will be caused by the economic push of increasing costs of the present system and the attractive pull of the new features that electronic publishing offers (Odlyzko,1996, p. 96)."

A similar prediction, which partially inspired the title of this article is articulated by Ginsparg (1996):

> "The essential question at this point is not *whether* the scientific research literature will migrate to fully electronic dissemination, but rather *how quickly* this transition will take place now that all of the requisite tools are on-line."

The **convergence** argument is more subtle, and is often implicit in the arguments of electronic publishing advocates.  The implication is clearest when e-publishing advocates focus upon only one or two kinds of forums (such as a pure e-journal or a working article server, like http://www.arxiv.org/) and do not explicitly discuss the heterogeneity of  disciplinary publication practices.

The electronic scholarly publishing reform movement has played a major role in encouraging scientists to take electronic media seriously as an opportunity for communication that may be faster, more accessible or even less costly than traditional alternatives. But it also becomes a source of tensions when enthusiasts (implicitly)  claim that a single model for electronic scholarly publishing is appropriate for all scholarly communities ("One size fits all").  A few electronic publishing enthusiasts dominate the discourse about the subject, and thus set expectations for the potential of electronic e-media in scientific communication.

What is often left out of the mainstream discussion is a more varied and subtle set of experiments by scientists who have developed electronic media, such as electronic journals, that fit the specifics of their discipline and that use electronic media in novel ways.  These more modest projects are not touted as universal solutions that most fields can emulate.

For example, Holoviak &  Seitter (1997) characterize their approach to  *Earth Interactions*, a pure e-journal for earth sciences:

> "From the beginning, we were determined to have a journal that would do much more than reproduce electronically what could be printed on the page. Our goal has been to exploit the medium and go beyond the capabilities of the printed page. *Earth Interactions* authors are encouraged to include sophisticated graphics, data in electronic formats, and even useable computer code -- the same tools they use to draw their conclusions. Readers can interact with parts of the author's data and observations and thus more readily verify and expand  the results presented."

In contrast, the editors and organizers of *the Electronic Transactions on Artificial Intelligence*(ETAI) sought to make the review process of articles more open for authors and readers (http://www.ida.liu.se/ext/etai/).

The *ETAI* represents a novel approach to electronic publishing. We do not simply inherit the patterns from the older technology, but instead we have rethought the structure of scientific communication in order to  make the best possible use of international computer networks as well as electronic document and database technologies.

---

[3]        It is interesting to note that both Harnad's "subversive proposal" and Ginsparg's E-Print server bypass peer-review (although Harnad also values peer-review and discusses a way of augmenting his "subversive proposal" to include peer-review).

Articles submitted to the *ETAI* are reviewed in a 2-phase process. After submission, an article is open to public online discussion in the area's News Journal. After the discussion period of three months, and after the authors have had a chance to revise it, the article is reviewed for acceptance by the *ETAI*, using confidential peer review and journal level quality criteria. This second phase is expected to be rather short because of the preceding discussion and possible revision. During the entire reviewing process, the article is already published in a "First Publication Archive", which compares to publication as a departmental tech report. (From ETAI, 1997; See Sandewall (1998) for a more elaborate description of their editorial process.)

The *ETAI* is an interesting contrast with another hybrid paper-electronic journal for artificial intelligence , the *Journal of Artificial Intelligence Research (JAIR)*.(See http://www.jair.org/). Articles in the *JAIR* are similar in content, format and referee processes to articles in paper journals, although there can be on-line appendices and discussions of published articles. People are encouraged to cite *JAIR*'s articles as they would articles in a paper journal. However, *JAIR* is distributed without charge on the Internet. (See Kling & Covi, 1995 for a more complete account)[4]. Each of these artificial intelligence e-journals values peer review, and public discussion, but also structures them a bit differently.

There are numerous other electronic journals whose editors are experimenting with new formats. These experiments are often tailored to fit the preferences of a specific disciplinary community and are relatively invisible despite their creative ingenuities. Our main point is **not** that a working article server used by high-energy physicists is of lesser value than peer-reviewed electronic journals. Rather, as we shall discuss, these different forums are structured in ways that are more fine tuned to the communicative practices and interests of different disciplines. This heterogeneity should be analytically understood when we characterize viable approaches to scientific communication.

## Social Shaping of Communication Technologies

There are many possible perspectives through which to analyze these manifest between-field differences in the scientific communications systems. Each perspective has its own particular set of phenomena that it throws focus upon, and its own set of phenomena that are ignored, or are treated as exogenous, or uninteresting.

One set of perspectives is typically used by electronic communication and publishing enthusiasts. From an information-processing perspective – a perspective that considers only the technical features of the various media -- all of these forums– pure electronic journals, databases, preprint servers, and pe-journals, are essentially equally valuable in all disciplines. They all are said to reduce the costs of communication, expand the range of people and locations from which materials are accessible, and generally speed communications. As scholars in all scientific fields work with data, and communicate both formally and informally with other scholars, all of these electronic media forums should be adopted and used fairly uniformly.

Fields do differ in some obvious ways based on the work products of the field. For example, computer-scientists don't generally work with shared, static-but-growing datasets, while molecular biologists frequently do. Therefore, it is not surprising that molecular biologists work with a number of shared digital disciplinary corpora, while computer-scientists do not. On the other hand, all scholars communicate formally. While pure electronic journals have been established in some areas of computer-science such as AI and mathematical theory, they play a negligible role in the communication systems of many other fields.

Another similar perspective is an evolutionary one: various fields, through somewhat random experimentation, have developed a series of electronic communications forums. Particle physicists have simply stumbled upon the concept of working paper (e-print) servers, and developed a dozen prototypes. Computer scientists have stumbled upon another architecture for organizing access to their working papers via the Networked Computer Science Technical Reference Library – NCSTRL (http://www.ncstrl.org) -- a distributed digital library of technical reports from major computer science departments. Many experiments simply fail, or never get off the ground. For example, the *Internet Journal of Science: Biological Chemistry*, which published four articles in 1997, and no articles in 1998. Soon, we should expect scientists in all fields to adapt these successful discoveries to enhance their communications. Thus, this perspective would hold that it is simply a matter of time (perhaps simply a matter of waiting for today's Internet-savvy students to become working scientists) before scientists of various fields adopt various communications innovations. A similar perspective, referred to earlier in this article as the *inescapable imperative*, views the adoption of various electronic communications forums in science as simply a small part of a much larger technologizing force sweeping over society. In this view, certain scientists, such as high-energy physicists, may simply be on the leading edge of an inexorable trend, and it is only a matter of time before other sciences will catch up with the avant-garde physicists.

A socially informed perspective, however, that we find particularly useful, is what has been called *social shaping of technology (SST)*. SST (MacKenzie & Wajcman, 1985) is a broad theoretical stance that views technologies not as autonomous causal agents driven by an internal, features-based technical logic, but rather as products of human creation and use. The configuration of socio-technological ensembles is driven by a series of operational choices (conscious and unconscious) made both

---

during the creation and during the implementation of the technology. A key insight from SST is that the configurable nature of many technologies – particularly IT – enables this shaping to continue during their use (Silverstone & Hirsch, 1992; Williams, 1997). The social shaping perspective is not inconsistent with a diffusion perspective; rather, the two perspectives focus attention on different phenomena. Diffusion of innovation, for example, would consider new electronic communications forums as exogenous, their success or lack of success determined by the "fit" of the technology with an existing work style of a community (Rogers, 1995). If the innovation does not fit sufficiently with the work style, and does not meet other criteria, it stagnates or dies. A social shaping perspective, however, focuses on the ongoing dynamic between a technology and a community, as the technology is developed, used, shaped, reconfigured, and reconstituted within the community.

We also draw many sensitizing concepts from another theoretical perspective – New Institutionalism. New Institutionalism is an essentially cognitive approach to organizational behavior, which roots in a variety of disciplines, including sociology, economics, and organizational studies, that posits that many choices of organizational actors are governed by highly routinized habits, scripts, rote actions, and imitation of elites; these routinized actions are conditioned by and reinforced by centralized institutions (DiMaggio & Powell, 1991). New Institutionalism is fully consistent with a social shaping approach; institutions are one important social force that shapes behavior (usually serving as a source of inertia, rather than radical change). One particular sensitizing concept that we can draw from New Institutionalism, and that we will return to later in this paper, is the importance of **legitimacy** or perceived legitimacy in determining action.

In the following analysis, we will adopt a social shaping perspective, sensitive to institutional forces, in order to analyze field differences in the use of electronic media.

## Social Shaping of Configurable Technologies

Williams (1997) posits a tension between two sets of opposing social forces: those that tend to stabilize technologies and those that tend to destabilize them. The destabilizing forces emerge both from advocates of technological innovation and from the changing needs of diverse groups of participants. Further, the increasing configurability of information and communications technologies (as opposed to, for example, mechanical industrial technologies) widens the potential for shaping IT in use. This insight from the SST position is consistent with our observations that a highly configurable technology such as the World Wide Web can be adopted and used by different fields in dramatically different ways. The implication is that the shaping of technology is highly specific to and emerges in reaction to the dynamic needs of specific communities, such as the active participants in scientific fields.

New IT organized with computer and telecommunications systems enables some people to communicate more rapidly, for a wider variety locations, and sometimes at low marginal costs. This value -- enhanced communication -- is one value that scientists value. It's important to note, though, that people's preferences about who to communicate with whom about what is highly contextualized. Scientists who may wish to have their published articles widely and rapidly read may also be reluctant to have drafts of their manuscripts or notes about new projects similarly available for a larger community.

Therefore, enhanced communication is not an absolute positive value in scientific fields. Scientists often want know about recent findings and new theoretical developments as it may effect their own research strategies and practices. Scientists are also concerned with receiving credit for their contributions, and to have such credit reflected in their status, career prospects, and abilities to obtain research resources and good jobs. These values can conflict in different ways in different scientific fields.

## Electronic Publishing and Scholarly Societies

Scholarly societies play a major role in the shaping of communications forums within a field, both because they are typically major publishers within a field, and also because they articulate and disseminate research and publishing standards for a field. There is a wide discrepancy between various scientific societies in their stances towards the "posting of documents on the Internet". These stances are reflected most clearly in the prior publication policies and practices of these societies – that is, to what degree is some form of "posting a document on the Internet" (Kling and McKim, 1999b) treated as "prior publication" by the editor of society journals.

One of the most widely publicized Internet publication policies came from the American Psychological Association (1996) whose "interim policy" asserted:

"Authors are instructed not to put their manuscripts on the Internet at any stage (draft, submitted for publication, in press, or published). Authors should be aware that they run a risk of having (a) their papers stolen, altered, or distributed without their permission and, very importantly, (b) an editor regard such papers as previously "published" and not eligible as a submission-a position taken by most APA journal editors. In addition, after acceptance for publication, the publisher is the copyright holder. APA does not permit authors to post the full text of their APA-published papers on the Internet at this time, as developments in the online world cannot be predicted. The APA will, however, closely follow such Internet developments. The P&C Board will establish a task force in June 1997 to investigate developments and recommend a longer term APA policy."

The APA's policy has been modified to devolve the decisions about Internet publishing to psychology journal editors. But the APA insists on its original policy for its own journals, which serve as major journals in the field. Lest the APA embargo on Internet posting seem anomalous, the American Chemical Society (ACS) has a similar policy for the *Journal of the American Chemical Society*:

> "As stated in the Notice to Authors of Papers submission of a manuscript to the Journal implies that the work reported therein has not received prior publication and is not under consideration for publication elsewhere in any medium, including electronic journals and computer data bases of a public nature. The editors and the advisory board have established a policy that any material that is posted in electronic conferences or on WWW pages or in newsgroups will be considered as published in that form, in the same way as if that work had been submitted or published in a print medium (American Chemical Society, 1996)."

These policies stand in contrast with the practices of some other fields, such as computer science and particle physics. The Association for Computing Machinery (ACM)'s interim copyright policy, for example, does not homogenize all forms of posting on the Internet, nor does it declare the posting of a document at any stage of development as equivalent to publishing. In part, the ACM's Interim Copyright Policy states that:

> "ACM intends to be the author's agent in reaching the widest possible readership and protecting the author's interests against plagiarism and unauthorized copying or attribution of an author's work. The ACM grants authors liberal reprint rights including unlimited reuse of the work with citation of the ACM publication and the right to post preprints and revisions on a personal server (ACM, 1995)."

ACM sees its role not as the sole provider of a work, but rather as a facilitator of wide readership access, and maintainer of the "version of record" of the author's work. Only the "definitive", published article need be maintained on the ACM Web server.

The ACM policy parallels the practices found in the particle physics community as well. When an author submits a article to a journal for publication in particle physics, the author typically posts the document *at the time of submission* on one of many publicly available working paper (or "e-print") servers. This document is then available for others to read, even before it has been received or reviewed by the journal.

*MIS Quarterly*, a high impact journal in the Information Systems field, represents yet a similar practice. Authors of articles to be published in *MIS Quarterly* may post article drafts on their own Web pages, with explicit notice that these drafts are "pre-prints", and are thus not the official, "published" versions of the article. Links to these drafts are collected on a Forthcoming Articles page on the *MIS Quarterly* Web site. However, once the journal issue is available, the author's draft must be taken down from the author's personal Web page (*MIS Quarterly* Web Site, 1998).

## The Role of Trustworthy Communication in Conceptualizing Field Differences

We can also characterize the basis for field differences in electronic media in terms of each field's articulation of some key issues faced by all scholarly fields: First, the allocation of credit for work performed; second, selection of target audiences for research; third, access to resources, including data; fourth, speed of work and results-sharing; fifth, allocation of professional status. The manner in which each field deals with these issues are both socially shaped and strongly institutionalized.

**Trust** plays a central role in the articulation of these issues by different fields, in two different senses.

(1) For scientists to be willing to read or use a report, they must be willing to trust that the report is legitimate (and that the study reported is competently executed and worth the time to learn about). Formal peer-review is only one of many processes of legitimization. Particle physicists and astrophysicists, for example, are more willing to use working papers than are molecular biologists or sociologists.

(2) Conversely, scholars who are willing to share materials (data, working papers, research reports, etc.) must have enough confidence that the sharing will not hurt their own career advancement or future access to resources. For example, if a scientist publishes an article in a pure electronic journal, will this be considered a "wasted publication" from a career-advancement perspective? If a scientist posts a working paper in advance of publication (or even acceptance for journal publication, as is currently done by many physicists), are they taking a risk that someone else will either (a) plagiarize their work, or, more commonly (b) take their work further, more quickly, and produce a higher-impact report?

Researchers in some fields, such as high energy physics and computer science have been quite willing to circulate their working papers quite widely, in paper formats and now via electronic media. In contrast, many chemists, molecular biologists and psychologists prefer to read peer reviewed articles when the topic is outside the envelope of their current expertise.

We have spent considerable time puzzling over the differences in communicative practices in different fields – and especially the extent to which researchers are willing to publicly share working papers prior to their acceptance by peer reviewed journals. We believe that the social conditions that lead scholars to shape the significance of peer-reviewing and formal publication in their communication systems are influenced by four important (and overlapping) structural characteristics of their fields[5].

## Research Project Costs

The overall costs of a project may have several effects on the communications system of a field. First, it may tend to increase collaboration, as it becomes difficult for any one researcher to mobilize the resources necessary to perform the research. Second, it may increase visibility of the work. Third, many specialized extramurally funded research institutes establish stronger ongoing controls for publishing research results, even as working papers. Institutes as diverse as CERN and the RAND Corporation are known for their internal reviews. High cost (multi-million dollar) research projects usually involve large scientific teams who may also subject their research reports to strong internal reviews, before publishing. Thus a research report of an experimental high-energy physics collaboration may have been read and reviewed by dozens of internal reviewers before it is made public.

## Mutual Visibility of Ongoing Work in the Field

Our studies indicate that productive scholars in some fields are more aware of the work that others in the field are doing than those in other fields. For example, most productive physicists are well-aware of the large-scale, high-cost collaborations in their field, and well in advance of its eventual publication in a journal. In contrast, few sociologists are aware of the ongoing work of other sociologists beyond their own close colleagues until they read about the research in a journal. If the ongoing work in a field is relatively transparent to others in the field, it may be that the risks associated with sharing reports and data may be less. This may also be the case in very small fields that publish almost exclusively in specialty outlets.

## Degree of Industrial Integration

Industrial collaborations, especially those that may readily result in income from patents and trade secrets, puts pressure on academics to be more conservative about sharing data and to sometimes delay publication about their methods and results.

## Degree of Concentration of Communication Channels (especially journals)

In some fields, a few journals serve as outlets for the vast majority of high quality published research while other fields routinely spawn specialized new journals. For example, astrophysics relies upon 3-4 major journals worldwide, while neurology relies upon well over 100. Research studies that are published in a few journals that are read by most scholars is much more likely to be visible than research that may be scattered through numerous journals.

## Translations from Paper to Electronic Media

Scientists have worked out ways of resolving issues of legitimacy and trust in traditional media. As a consequence, when scientists computerize traditional media, they are readily organized in ways that protect traditional concerns. Thus, for example, scientists are more willing to accept the electronic versions of paper journals (such as *Science*), than to accept new pure electronic journals (Kling & Covi, 1995). Similarly, in the 1960s, high energy physicists had clearinghouses for sharing working papers. Electronic versions, such as the Los Alamos National Labs E-Print Server and the CERN preprint server are relatively uncontroversial extensions of these services. In *Drosophila* research, a book of DNA sequences (The *Red Book*) was developed in the 1960s. FlyBase, its electronic version, was also non-controversial This pattern of translating traditional forms in electronic media is predicted by institutional theories.

It is important to note that many of these electronic translations are not simple replications in electronic form of what was previously made available in print form. They can include numerous value added features, and capture some of the gains of IT in more rapid communication among specialists in the field. One important family of practices is occurring with **p-e journals** – journals that circulate primarily as paper journals but that also support electronic versions. P-e journals have become the loci of a wide variety of electronic enhancements. Some p-e journals, such as *The Astronomical Journal*, are posting searchable abstracts and the full text of articles on-line. (Anyone can read the abstracts, but the full texts is available only to subscribers). The *Astronomical Journal* also posts lists of articles that are under review, along with contact information for those who wish to contact the authors for preprints (see http://www.astro.washington.edu/astroj/lemon/month.html). The *Journal of the American Chemical Society* posts the current issue online, and also maintains a publicly readable on-line archive of a few selected "hot articles" (see http://pubs.acs.org/journals/jacsat/index.html). *Science* provides links to *Medline* abstracts of cited articles as well as links to supplementary data sets. These examples illustrate just a few of the numerous efforts of paper-journal publishers to create elaborate electronic value-added enhancements, while they preserve existing editorial control over web posting and peer review (see Kling & McKim, 1999b).

Of course not all translations from paper to electronic media survive. For example, some molecular biologists who study the worm *c.elegans* created an online Worm Community System (WCS). It proved to relatively complicated and made many technical demands upon its users (Star & Ruhleder, 1996). It was abandoned, but was replaced by a web-based system , ACEDB – "A C. Elegans DataBase" (http://probe.nalusda.gov:8000/other/aboutacedb.html) that was much more workable, in terms of fitting the technological work routines of biological researchers. ACEDB originally replaced the WCS on CD-ROM and then was made available as a WWW-based resource.

---

[5] We have identified these structural characteristics from our previous research, including interviews with scientists and scholars in about 10 disciplines, and are currently engaged in a larger study that will clarify and test the characteristics.

The American Association for the Advancement of Science spent about $2 million to develop *an Online Journal of Current Clinical Trials*, and sold it for 15% of that cost several years later. It now appears to be inactive. Many biologists invested in the Genome Database (GDB), a database that supports the mapping phase of the Human Genome Project (Letovsky, 1998. However, its financial support had been withdrawn because its funders do not see adequate value for their constituencies[6]. These are just a two high profile examples.

Over time, say the next century, we should expect many new kinds of electronic forums to support scientific communication. It is not hard to find de novo creations today, such as pure e-journals or shared disciplinary compendia, such as ISWORLD (see above) and GDB. There is a risk that these innovative e-media may not thrive nor survive. While ISWORLD seems successful today (based on a growth of contributions), GDB narrowly escaped being closed down. In addition, many pure e-journals struggle to attract high quality articles and an effective readership.

It is likely that some conventions of scientific communication will change in the next decades, especially when they can be organized to preserve or enhance trust and bring other benefits as well. We suspect that fields will differ in the ways that they shape e-media because trust issues work out differently with respect to characteristics such as those that we have identified above (i.e. visibility of projects to others in the field, industrial integration).

## Conclusions

This is a relatively early period in the development of electronic media to support scientific communication. It's important to view many of today's adventures in electronic publishing as experiments. Since 1990, scientists, publishers and librarians have initiated thousands of electronic publishing ventures, include new electronic journals and disciplinary databases.

Some of these have thrived, while others have not developed sustainable support. Most of these systems may have seemed highly promising when they were initiated. Short life spans are not unique to electronic journals and on-line databases. Some fraction of paper journals are also disestablished after a few years if they don't "catch on" with authors, readers and subscribers. But we should do our best to capture the benefits of electronic media without paying unnecessarily high experimental costs or "losing scientific reports" that are published in what soon become disestablished and little visited electronic islands.

We do not criticize experiments in electronic publishing per se, and are optimistic about the potential for significant enhancement of scholarly communication through the use of electronic media. What we have critiqued in this paper is the information processing theorizing about electronic publishing and scholarly communication that dominates both popular and academic discourse about the subject – theorizing that elides and homogenizes field differences.

One key principle for good scientific theories of systems developments is that they should be conceptually rich enough to *understand or predict variations in behavior and structures, such as failures as well as successes*. The information-processing theories that fuels some of the most strident evangelism about electronic publication do not satisfy this criterion. The history of discourses about appropriate forms of computerization is littered with utopian visions that do not effectively engage the complexities of the social worlds of the likely users of new technologies (Kling & Lamb, 1996).

Our institutional social shaping approach examines electronic media comparatively and as shaped by the social practices that support trustworthy communications in diverse scientific disciplines. This approach is media-independent, empirically accessible, and gives some promise of helping us better understand what it takes for scientists to organize electronic media to effectively energize their communications in practice.

The future of electronic publishing in the next few decades will likely be enacted through institutional incrementalism. For example, a number of scientific societies and trade publishers have developed varied forms of electronic enhancements to their print journals – including on-line abstracts, tables of contents, full text, supplementary data sets, and searchable text. It is these p-e journals that are really growing most strongly, both in number and in elaboration, when commentators refer to vast growth in electronic scholarly journals (e.g. Hitchcock, et al., 1996). We anticipate increased experimentation with and addition to these value-added electronic features. However, the primary business model for full access to these electronic enhancements of paper journals is based on membership dues and/or subscription fees. In addition – and most important -- these electronic value-added features are not turning peer-reviewed journals into unreviewed compilations.

We expect that considerations of trust will continue to shape the kinds of scholarly communication that are seen as legitimate in a specific field. The divide between fields where researchers share unrefereed articles papers quite freely ("open flow fields") and those where peer reviewing creates a kind of chastity belt ("restricted flow fields") is likely to change slowly, if at all. Thus, for example, Ginsparg's unrefereed and (largely) unrestricted working article server includes some areas of physics, and a few cognate mathematical and chemical subfields (with relatively few mathematics articles posted). But we expect few biological or chemical specialties to join forces with this venture, and embrace it like high energy physics or computer science. Even the field of physical chemistry has only a few papers posted in the first three months of 1999 (in contrast with experimental high energy physics with nearly 100 submissions).

---

[6] In December of 1998, the Bioinformatics Centre at the Hospital for Sick Children in Toronto agreed to take over funding for curation and data collection activities for GDB, rescuing GDB from shut down (Cuticcia, Talbot, Porter, Uberbacher, & Snoddy, 1998).

Currently, the National Institutes of Health (NIH) is funding and developing a Web-based electronic working paper server (to be called "PubMed Central") for all of biomedicine, (Marshall, 1999; National Institutes of Health, 1999). The original proposal for the archive called for a working paper server, modeled after Arxiv.org, that would have allowed concurrent access both to refereed and unrefereed articles deposited in the server by authors (Varmus, 1999). However, after extensive comment on the proposal by the biomedical community, and lobbying by scholarly societies, the proposal was significantly modified, and now emphasizes free (to the reader) access to refereed articles that have been deposited in the server not by the author but by a sponsoring journal or scholarly society. It is "not just a matter of time" for biologists to adopt the communication practices of high energy physicists. For a detailed study of this case, see Kling, Fortuna, King and McKim (forthcoming). There are a number of additional disciplinary author-posted on-line "working paper" archives, but they still cover only a few fields, such as mathematics and demography, and have not been as comprehensive in their coverage as has arXiv.org (Van de Sompel. and Lagoze, 2000).

On the other hand, we expect to see a cornucopia of experiments with new value-added features on either side of the refereeing divide: in particular, many paper journals will become p-e journals. And p-e journals are likely to add many valued added electronic enhancements for their readers. In addition, some new peer-reviewed pure e-journals, such as *ETAI* and *the Journal of the Association of Information Systems* may become accepted as high quality society-sponsored journals. Similarly, conferences will be continuing to add electronic enhancements, such as Web sites that are steadily augmented with program information, author contacts, and abstracts, and, in some fields, articles, after the conference is over. We expect that these various ventures will significantly enrich scholarly communication, even if they are not based on the free of charges instant-access model that some e-publishing reformers advocate.

## Acknowledgments

Funding was provided in part by NSF Grant #SBR-9872961 and with support from the School of Library and Information Science at Indiana University. Joanna Fortuna played a major role in developing Appendix A. This article benefited from helpful discussions with and comments from Lisa Covi, Blaise Cronin, Jeff Hart, Mitzi Lewison, Howard Rosenbaum, Erik Sandewall, Steve Sawyer, John Walsh and Luc Wilkin. In addition, we are indebted to numerous scientists who have discussed their publication and reading practices with us. An early version of this analysis was outlined in Kling & McKim (1999a).

## Appendix A: Illustrative Electronic Communications Forums

This Appendix briefly describes the forums that are identified in Table #1.

**Arxiv.Org: The Los Alamos National Laboratory e-print archive http://www.arxiv.org/ :** Founded in 1991 as an archive of full-text High Energy Physics working papers . Authors post their own articles (usually at the same time as, and in parallel with journal submission). There is little editorial control and no formal review. This freely accessible searchable archive now covers 8 major areas of physics and 5 related disciplines. By mid-1999 the number of accumulated submissions to the archive reached 100,000. Additional features include email lists through which new submissions are announced to interested subscribers, and threaded jobs and conference announcement lists.

**The Astrophysical Journal http://www.journals.uchicago.edu/ApJ/ :** Begun in 1895, The Astrophysical Journal is a peer-reviewed research journal devoted to recent developments, discoveries, and theories in astronomy and astrophysics, and is published 3 times per month for The American Astronomical Society (AAS). ApJ publishes peer reviewed research articles, letters (shorter, less rigorous and more speculative), and supplementary materials in print and electronic editions. The ApJ electronic edition currently contains almost 6,000 full-text articles, letters and supplements (1997 to the present). Older issues are available through the NASA Astrophysics Data System (ADS) service, which contains over 55,000 citations from ApJ. Authors are encouraged to submit articles electronically. Searching and full-texts are limited to authorized users (including over 700 institutional subscribers). Additional features: video supplements for some issues, list of all the articles recently submitted to ApJ, and abstracts of ApJ letters accepted but not yet published.

**The Astrophysics Data System (ADS) http://adswww.harvard.edu/ :** an abstract service which includes four searchable sets of abstracts dating back to 1975: 1) over 485,000 astronomy and astrophysics abstracts; 2) almost 514,000 instrumentation abstracts; 3) almost 430,000 physics and geophysics abstracts; and 4) almost 3,000 abstracts from the Los Alamos preprint server (www.arxiv.org). Abstracts come from hundreds of journals, publications, colloquia, symposia, proceedings, NASA reports and many PhD theses. Until 1995 the primary source of abstracts was NASA's STI database. Since then, the majority of the abstracts have been received directly from the journal editors. Authors, institutions and libraries can also submit abstracts to be included in the Abstract database. The database was recently supplemented with free scanned images of full-text articles of older issues.

**Chicago Journal of Theoretical Computer Science (CJTCS) http://www.cs.uchicago.edu/publications/cjtcs/ :** a peer-reviewed scholarly pure e-journal in theoretical computer science, published by MIT Press. Articles are submitted via e-mail and published on the Internet one article at a time. From its founding in 1995 to mid-1999 CJTCS has published between 4 and 6 articles a year (23 articles total). CJTCS experimented with restricting full text access to Internet domains of subscribing institutions, but returned to a policy of free full-text access. Paid subscribers receive an annual print edition and guaranteed archival access. In 1998 there were over 60 institutional subscribers to CJTCS.

**Computer Science journals published by Elsevier http://www.elsevier.com :**

Elsevier publishes over 50 peer-reviewed Computer Science journals. Institutional subscribers receive paper issues as well as electronic access to full-texts for 9 months after issue publication, and members of subscribing institutions can sign up to receive tables of contents or abstracts via e-mail several weeks prior to publication. Since 1995 Elsevier offers full-text electronic archive subscriptions to libraries through a service called ScienceDirect. Free public access to citations and abstracts search, and sample copies online with registration. Online services vary substantially from journal to journal. Some journals post full-texts of accepted but not yet published articles (for subscribers only).

**ACM Digital Library http://www.acm.org/dl/ :** Launched in 1997, by mid-1999 the ACM Digital Library is a searchable collection of citations and full-text of articles from 16 ACM journals, 6 ACM magazines, and conference proceedings from 57 distinct ACM conferences (published in over 400 volumes), and a few non-ACM journals and magazines. By 1998 the Digital Library included nearly 9,000 full-text articles from ACM journals, magazines, and conference proceedings from 1991 forward, and tables of contents with nearly 20,000 citations from 1985 forward. As of the end of June 1998, there were approximately 15,000 individual subscribers. Only ACM members who pay for the Digital Library subscription have access to all full-text articles. Access to tables of contents, citations, and abstracts is open to the public at no cost.

**ACM Journals http://www.acm.org/pubs/journals.html :** ACM offers over two dozen publications, including over 16 peer-reviewed scholarly journals, available in both paper and electronic formats. The full-text articles, abstracts and searching are available as part of the ACM Digital Library, but one can subscribe to individual journals electronic access. Full-text archives are usually available for recent years, and citations and abstracts for earlier issues. Some journals post articles submitted electronically after acceptance (before publication).

**Artificial                                       Intelligence**
http://www.elsevier.com/locate/artint : Founded in 1970, as the first specialized journal for artificial intelligence research, it became the top ranked journal in the field long before the electronic enhancements. In its first decade of operations, only 10 to 20 articles were published yearly; in the 190s between 60 and 120 articles are being published each year. The journal started out as a quarterly, but switched to a monthly publication system. Artificial Intelligence is one of the enhanced paper-and-electronic journals published by Elsevier. Online tables of contents of all issues are free. As of mid-1999 subscribers can search and browse over 1000 abstracts (1970 to the present) and over 200 full-texts of articles (1997 to the present). Additional features: electronic versions of accepted papers posted before publication; users can sign up to receive the abstracts of new issues by email upon publication.

**Electronic Transactions on Artificial Intelligence (ETAI)**
http://www.ida.liu.se/ext/etai/ : a peer-reviewed pure e-journal founded in 1997. After submission, articles are open to 3 months public attributed online preview. After the open review period, the author may revise the article, and only then is the manuscript peer-reviewed confidentially. During the entire reviewing process, a version of the article is available online. The electronic issues are free. An annual print edition is available for free to some selected institutions and researchers; a limited additional supply can be purchased by others. ETAI accepted 5 articles for publication in 1997 and 10 articles in 1998. In mid-1999 5 articles were under public review in ETAI and 1 article under confidential peer review.

**Fly Base    http://flybase.bio.indiana.edu/ :** aims to be a comprehensive database for information on the genetics and molecular biology of Drosophila melanogaster, the fruit fly. Evolved from a series of earlier print publications called the Redbook, last published in 1992. Includes a variety of data about all known gene sequences of Drosophila that have been curated through Western scientific literature so far: more than 51,500 variations (alleles) of more than 11,000 genes, more than 17,400 Drosophila stocks (live groups of flies) held in stock centers and private labs, a bibliography of over 93,000 Drosophila citations, an address book of over 5,800 Drosophila researchers, and references to about a dozen other categories of many 1000s of entries of other biological data. All data is extensively cross-linked. FlyBase also includes a wide variety of other materials including conferences, Internet discussion group archives, and information on and cross-links to several other closely related projects. The vast majority of information in FlyBase is collected from scientific literature or genome sequencing projects, though some data comes as personal communications from individuals. Includes data files, documents, indices, forms, and images. All material is accessible at no cost. Limited extracts of FlyBase data are periodically published as special issues of the periodical Drosophila Information Service (DIS).

**GenBank**
http://www.ncbi.nlm.nih.gov/Web/Genbank/index.html : a major data resource for genetic research which provides public access to a variety of genetic sequence databases, annotated with links to scientific article abstracts, different search engines and many other tools. The collection contains all known and publicly available DNA sequences of almost 30 different organisms. As of May 1999 GenBank included complete genomes (complete sets of genetic information) for 21 organisms and several additional DNA sequences and chromosome maps (including humans and fruit flies). GenBank acquires the DNA sequences primarily through direct submission of sequence data from individual laboratories and from large-scale sequencing projects. There are approximately 2,57 billion bases in 3,5 million sequence records. Many scientific journals require submission of genetic sequence information to a standard database prior to publication so that an accession number may appear in the paper. Based on genetic sequencing database projects established in 1979, GenBank now exchanges data daily with European and Japanese projects.

**The Genome Database (GDB)   http://www.gdb.org/ :** the official central repository for genomic mapping data resulting from the Human Genome Initiative, a worldwide research effort to analyze the structure of human DNA and determine the location and sequence of the estimated 100,000 human genes. Established in 1990 at Johns Hopkins University, GDB was almost terminated in 1998 due to funding problems. It was rescued at the last minute when the Hospital for Sick Children in Toronto, Canada agreed to provide support for the project. By mid-1999 GDB included over 45,000 genes, over 3,000,000 clones (groups of cells derived from a single ancestor), almost 85,000 amplimers (short chains of subunits of DNA used as markers in a PCR amplification reaction -- a method to amplify DNA), almost 800 maps and over 300,000 map elements. GDB also contains over 66,000 citations of supporting journal articles and personal communications, and contact information of almost 14,000 individuals, labs, commercial organizations, projects, and repositories. Human gene entries in GDB are linked to homologous genes (genes that play similar roles) in fruit flies through the FlyBase database. GDB was an early Web-based compendium in molecular biology. Although the database focuses on gene mapping, GDB mananagers plan to broaden the focus of GDB as the Human Genome Initiative moves from mapping to sequencing of genes. Members of the genomic research community can request an account on GDB, after which they can submit their data. Access to the database is free.

**Hubble Space Telescope (HST) Archive**
http://archive.stsci.edu/ **:** one of 3 archives of astronomical data and technical information about the Hubble Space Telescope (launched in 1990 and scheduled to operate through 2010). This is the primary archive and distribution center for HST data, distributing raw science data, calibration, and engineering data over the Internet to astronomical researchers. The archive catalog contains information about all observations that have been made with HST (over 100,000 observations of more than 20,000 targets; total size of approx. 6.34 Tbytes, 100-200 Gbytes added/month, 300-600 Gbytes retrieved/month). Use of the archive requires authorization as well as specialized software capabilities. Over 1,500 users are registered to use the archive (with limits and quotas); about 500 of these users are authorized to look at proprietary data from one or more of the approximately 1,000 proposals. The results of investigations with HST archive data are generally published in the scientific literature. The site also includes a special section written for the public.

**ISWorld** http://www.isworld.org/ **:** resources related to information systems (IS) for a worldwide community of ca. 5,000 researchers and practitioners, maintained by ca. 200 volunteers. Founded in 1994, the ISWorld Net site now includes working papers, course syllabi, instructional technology, research guidelines, annotated bibliographies, lectures, reviews, conferences and upcoming events, journal rankings, info on professional organizations and funding opportunities, a searchable directory of IS faculty and researchers worldwide, country-specific information for foreign researchers, etc. All materials on ISWorld are in public domain unless otherwise specified by the authors. Anyone can submit a page proposal. Although the material in ISWorld Net is not rigorously reviewed, it is solicited, loosely reviewed, organized, updated, and maintained. The site is supplemented by an e-mail discussion list.

**Journal of Artificial Intelligence Research (JAIR)** http://www.cs.washington.edu/research/jair/ **:** publishes refereed research articles, survey articles, expository articles and technical notes. Established in 1993 as one of the first electronic peer-reviewed scientific journals, during its first 5 years of operation JAIR evaluated nearly 600 submissions, publishing 114 articles. Articles are submitted by authors electronically, and published by JAIR online immediately upon acceptance. Subscription fees are required only to purchase the annual print volumes. The free electronic versions of the articles are already formatted and paginated as they would appear in a traditional printed journal (Kling & Covi, 1995). Some articles are supported by attached online code, data or demonstrationss. JAIR has its own Internet discussion group and distribution list. A project at the MIT Artificial Intelligence Lab designed an information space for searching JAIR articles arranged on a 3-D plane. (http://www.infoarch.ai.mit.edu/jair/jair-space.html).

**Journal of the Association for Information Systems (JAIS)** http://jais.aisnet.org/ **and Communications of the Association for Information Systems (CAIS)** http://cais.isworld.org/ **:** published by the Association for Information Systems (AIS) beginning in 1999. JAIS, a peer-reviewed e-journal, was scheduled to begin publication in mid-1999 but had not published anything by July 1999 yet. The entire publication process from submission, through review, revision, and final edit will be electronic. Articles will be put up immediately upon acceptance. CAIS publishes opinions and supplementary materials (such as detailed data, appendices, screen images, survey measures, and programs) – without page limits. As of mid-1999 CAIS had published 22 articles. Both journals will be delivered to subscribers electronically, and there will be no paper copy. All members of the AIS receive subscriptions to JAIS. Nonmembers of AIS and institutions can subscribe to the journal for an annual fee.

**Management Information Systems (MIS) Quarterly** http://www.misq.org/ **:** Founded in 1977as a peer-reviewed scholarly print journal. MISQ published 15 articles in its initial year of operation, and has been publishing a steady 20-25 articles per year ever since. The MISQ web site includes tables of contents and abstracts for all issues, past issues search capability, calls for papers, full-texts of non-peer reviewed editorial statements (since 1992), a list of (15 in mid-1999) forthcoming articles, links to non-final pre-publication versions of some of these accepted articles, free full text of some special papers, and appendices for selected MISQ papers, and tracking of the review process status of submitted manuscripts. MIS Quarterly is one of the founders of ISWorld Net, and founder of the electronic journal MISQ Discovery.

**MISQ Discovery** http://www.misq.org/discovery/home.html **:** Founded in 1995 and envisioned as a revolutionary new electronic publication forum unlike traditional paper journals (and yet peer-reviewed), MISQ Discovery aims to publish not only research reports, but also multimedia, video, interactivity, hypertext, and live data pertinent to Information Systems; not only static (archival) works, but also dynamic (living, updated) works. MISQ Discovery is open to free access by. Works are published individually upon acceptance, announced on ISWorld, and abstracted in MIS Quarterly (the two publications share the editorial board). Due to extremely low submission rates, MISQ Discovery has published only 3 articles so far (one each in 1996, 1997, and 1998).

**NCSTRL (pronounced "ancestral") Networked Computer Science Technical Reports Library** http://www.ncstrl.org/ **:** an international distributed online digital library collection of Computer Science technical reports, submitted by authors from 157 participating institutions. Started in 1995 (but based on preceding collections), NCSTRL currently includes over 30,000 tech reports. Searching and online access to full texts on NCSTRL is free to anyone; hardcopies can be requested for a fee; and stored search queries to be processed automatically on a periodic basis are available through a subscription service.

**Protein Data Bank (PDB)   http://www.rcsb.org/pdb/ :** the single international repository for public data on the 3-dimensional structures of biological macromolecules. The data bank contains more than 10,000 structures (including among others proteins, enzymes, nucleic acids, and carbohydrates). Researchers can deposit structures, process and validate data themselves. Deposited data are further processed by staff before being released. PDB enables scientists to perform simple analysis or to download structure files for further analysis. Software tools for 3D structure visualization are provided. Founded in 1971 at the Brookhaven National Laboratory (BNL), and funded by the National Science Foundation, operations of PDB were recently transferred in 1999 to the Research Collaboratory for Structural Bioinformatics (RCSB). The National Institutes of Health, as well as many journals, require submission of certain types of data at time of journal publication of experimental results.

**Science Online   http://www.scienceonline.org :** An enhanced version of the magazine/journal published by the American Association for the Advancement of Science (AAAS). Science, published since 1880, now circulates about 300,000 paper copies per weekly issue. Science publishes both magazine-style scientific news (written by editorial staff) and peer-reviewed scientific papers in the life and physical sciences (about 20/week). Started in 1996, Science Online currently contains over 180 back issues of Science. Features/Access: searchable and browsable full texts of Science papers and personalized research alerts via email (subscribers only); summarized contents (free after registration); tables of contents, supplementary data, audio-visual, questionnaires, online discussions, funding and job listing databases, and announcements of scientific meetings (free without registration).

**SLAC-SPIRES (Stanford Linear Accelerator - Stanford Public Information Retrieval System) HEP (High Energy Physics) Database.   http://www-spires.slac.stanford.edu/FIND/hep/ :** The HEP project at SPIRES-SLAC is a joint project of the SLAC and DESY Physics labs. Founded in 1974, this searchable database now includes almost 400,000 bibliographic records, representing preprints, e-prints, journal articles, technical reports, conference papers, theses, etc. – full texts of over 150,000

are available to the public for free. A staff of about 10 people select items for inclusion, add specialized keywords, and maintain the system.

**Springer Verlag – LINK http://www.link.springer.de/ol/csol/index.htm :** Since 1997 Springer Verlag, an international publishing group, offers a digital library called LINK, which is not restricted to Springer journals and is available for institutional or individual subscriptions. The journals' tables of contents and article abstracts can be read and a full-text search implemented without charge. Full-text articles and alerting services are available to LINK subscribers registered for the journal in question. LINK includes more than 45 computer science journals -- published in paper format only, paper and electronic full-text, electronic journal only, paper with electronic abstracts only, or with electronic supplementary materials. A few of the journals are published online first, weeks before distribution of the print journal in their final form without. LINK includes a mixture of new and established journals with varying archival availability.